\documentclass{ws-ijmpcs}

\def\trimmarks{}

\newcommand{\beq}{\begin{equation}}
\newcommand{\eeq}{\end{equation}}
\newcommand{\ba}{\begin{eqnarray}}
\newcommand{\ea}{\end{eqnarray}}

\newcommand{\cdott}{{\mskip -1.5mu} \cdot {\mskip -1.5mu}}
\newcommand{\ms}{\mskip 1.5mu}

\begin{document}

\markboth{Dani\"el Boer}
{Aspects of TMD evolution of azimuthal asymmetries}

%
\catchline{}{}{}{}{}
%

\title{\mbox{}\\[-10 mm]
ASPECTS OF TMD EVOLUTION OF AZIMUTHAL ASYMMETRIES\footnote{Invited talk given at the QCD Evolution Workshop (QCD2013), 
Thomas Jefferson National Accelerator Facility, Newport News, VA, May 6-10, 2013}}

\author{DANI\"EL BOER}

\address{Theory Group, KVI, University of Groningen\\
Zernikelaan 25, NL-9747 AA Groningen, The Netherlands\\
d.boer@rug.nl}

\maketitle


\begin{abstract}
In this contribution TMD evolution of azimuthal asymmetries, in particular of the Sivers and double Collins asymmetries, is addressed.  
A comparison of the scale dependence is made between asymmetries described with TMD factorization at low transverse momentum and those described with collinear factorization at high transverse momentum. Finally, the advantages of Bessel weighting are discussed: convergence of transverse momentum integrals, suppression of large transverse momentum contributions, and well-defined lattice QCD evaluations of Bessel-weighted TMDs including proper gauge links.  
\end{abstract}

\ccode{PACS numbers: 12.38.-t; 13.88.+e}

\section{Introduction}	
The Drell-Yan (DY) process (lepton pair production in hadron-hadron collisions), back-to-back hadron production in electron-positron annihilation ($e^+e^- \to h_1 h_2 X$), and semi-inclusive deep inelastic scattering (SIDIS) have in common that all three are multi-scale processes 
that involve a large or hard scale ($Q$), a transverse momentum scale ($Q_T$) and a hadronic scale ($M$). 
Generally these processes do not take place in one plane and azimuthal asymmetries arise. 
If $Q_T \ll Q$, such asymmetries are most 
naturally described within the framework of transverse momentum dependent parton distributions (TMDs). 
Several forms of TMD factorization have been considered in the literature 
[\refcite{CS81}-\refcite{EIS2012}] 
(also for additional processes [\refcite{SXY,MWZ}]), which mainly differ in the treatment of rapidity and light-cone 
divergences (in order to make each factor in the factorized expression well-defined) and in the distribution of contributions
among the various factors in order to avoid the appearance of large logarithms. TMD factorization 
dictates the energy evolution of azimuthal asymmetries. This was studied in the various TMD factorization approaches in 
[\refcite{B01}-\refcite{Boer:2013zca}]. TMD evolution leads to decreasing asymmetries with increasing energy. This dependence on 
the hard scale generally differs from that in the high transverse momentum region, complicating the determination of the scale dependence of observables 
partially integrated over transverse momentum or integrated with transverse momentum weights. 
 
\section{TMD evolution of azimuthal asymmetries}

The new TMD factorization proven by J. Collins in 2011 [\refcite{Collins:2011zzd}] is of the form:
\beq
d \sigma = {H \times \text{convolution of}\ A\, B} + \text{high-}q_T\ \text{correction}\ (Y)+ \text{power-suppressed}
\eeq
Here $A$ and $B$ are transverse momentum dependent parton distribution functions (TMD pdf's) or fragmentation functions (TMD FF's), 
which in transverse coordinate space are functions of 
$x$, $b_T$, a rapidity $\zeta$, and the renormalization scale $\mu$. $H$ is the partonic hard scattering factor. 
The soft factor appearing in earlier forms of TMD factorization has been absorbed into $A$ and $B$, so as to 
have a TMD definition that is free from rapidity and Wilson-line self-energy divergences. 
For a brief summary cf.\ [\refcite{Collins:2011ca}]. 
This form applies to $e^+e^- \to h_1 h_2 X$, SIDIS and DY. 

More specifically, the differential cross section for SIDIS is given by:
\beq
\frac{d\sigma}{dx dy dz d\phi {d^2 \bm{q}_{T}^{}}} = \int d^2 {b} 
\, e^{-i {\bm{b} \cdot \bm{q}_T^{}}} \tilde{W}({\bm{b}}, Q;
  x, y, z) + {\cal O}\left(Q_T^2/Q^2\right).
\eeq
For unpolarized hadrons and unpolarized quarks of flavor $a$ the integrand $\tilde W$ is
\beq
\tilde{W}({\bm{b}}, Q; x,y,z) =
{\sum_{a}\,\tilde{f}_1^{a}(x,{\bm{b}};\zeta_F, \mu)}
{\tilde{D}_1^{a}(z,{\bm{b}};\zeta_D,\mu)} H\left(y,Q;\mu\right).
\eeq
For the choice $\mu = Q$ the partonic hard scattering part $H$ is of the form $H\left(Q;\alpha_s(Q)\right) \propto  e_a^2 \left(1 + \alpha_s(Q^2) F_1 + {\cal O}(\alpha_s^2) \right)$,
where $F_1$ is a renormalization scheme dependent finite term. By using the Collins-Soper and renormalization group equations [\refcite{Collins:2011zzd}], one can 
express the TMDs at the fixed scale $Q_0$, which is taken to be the lowest scale for which perturbation theory is expected to be applicable (we take 
the standard choice $Q_0=1.6$ GeV and $\zeta_F=\zeta_D$):
\[
\tilde{f}_1^a(x,b^2; \zeta_F, \mu) \, 
\tilde{D}_1^b(z,b^2;\zeta_D, \mu) = e^{-S(b,Q,Q_0)} \tilde{f}_1^a(x,b^2; Q_0^2, Q_0) \, 
\tilde{D}_1^b(z,b^2;Q_0^2, Q_0).
\]
The perturbative expression for the Sudakov factor $S$ is:
\[
S_p(b,Q,Q_0) = \frac{C_F}{\pi}  \int_{Q_0^2}^{Q^2}  \frac{d\mu^2}{\mu^2} \alpha_s(\mu) \ln \frac{Q^2}{\mu^2} -  \frac{16}{33-2n_f} \ln\left(\frac{Q^2}{Q_0^2}\right) \ln \left[\frac{\ln\left(\mu_b^2/\Lambda^2\right)}{\ln\left(Q_0^2/\Lambda^2\right)} \right].
\]
Using only this $S_p$ for $S$ is valid for $Q^2$ very large, when the restriction $b^2 \ll 1/\Lambda^2$ is justified.
If also larger $b$ contributions are important, i.e.\ for non-asymptotic $Q$ and small $Q_T$, 
one needs to include a non-perturbative Sudakov factor $S_{NP}$, e.g.\ 
$\tilde{W}(b) \equiv \tilde{W}(b_*) \, e^{-{S_{NP}(b)}}$, 
where $b_*=b/\sqrt{1+b^2/b_{\max}^2} \leq b_{\max} \sim 1/Q_0$. $W(b_*)$ can be calculated within perturbation theory.
In general $S_{NP}$ is of the form ${S_{NP}(b,Q)} = {\ln(Q^2/Q_0^2)}{g_1(b)} + {g_A(x_A,b)} + {g_B(x_B,b)}$  [\refcite{Collins:1985kw,CSS-85}], 
but cannot be  calculated perturbatively. It should be fitted to 
data and is necessary to describe available data. Fits of $S_{NP}$ to DY and Z production data have been discussed in Refs.\ 
[\refcite{Ladinsky:1993zn}-\refcite{Konychev:2005iy}].
Recently, SIDIS data was included in [\refcite{AR}], leading for $x=0.1$ to $S_{NP}(b,Q,Q_0) = \left[0.184 \ln\frac{Q}{2Q_0}+0.332 \right] b^2$ (cf.\ [\refcite{Sun:2013dya}] for criticism). At this workshop J. Collins argued for non-Gaussian $S_{NP}$, e.g.\ $e^{-m\left(\sqrt{b^2+b_0^2}-b_0\right)}$, as did D. Sivers. 

\subsection{TMD evolution of the Sivers effect}

The analyzing power $A_{UT}$ of the Sivers $\sin(\phi_h-\phi_S)$ asymmetry in SIDIS is  
\beq
A_{UT}(x,z,Q_T) = \frac{x\, z^2\,\left(1-y+\frac{1}{2}y^2\right) \sum_{a,\bar a} e_a^2 \int db\,b^2 \,J_1(bQ_T) {\cal N}_a}{x\, z^2\, \left(1-y+\frac{1}{2}y^2\right)M Q_T  \sum_{b,\bar b} e_b^2 
\int db\, b \, J_0(bQ_T) {\cal D}_b}, 
\eeq
where the integrand factors in numerator and denominator are given by
\ba 
{\cal N}_a & = & \tilde{f}_{1T}^{\perp\prime \; a}(x,b_*^2;Q_0^2,Q_0)\, \tilde{D}_1^a(z,b_*^2;Q_0^2,Q_0) \exp\left({-S_p(b_*,Q,Q_0)-S_{NP}(b,Q/Q_0)}\right),  \nonumber \\
{\cal D}_b & = &  \tilde{f}_1^b(x,b_*^2;Q_0^2,Q_0) \tilde{D}_1^b(z,b_*^2;Q_0^2,Q_0)\exp\left({-S_p(b_*,Q,Q_0)-S_{NP}(b,Q/Q_0)}\right), \nonumber
\ea
and $-2i b^i \tilde{f}_{1T}^{\perp \prime \; a}(x,b^2) \equiv \int \frac{d^2 \bm{p}_{T}^{}}{(2\pi)^2} e^{i\bm{p}_{T}^{} \cdot \bm{b}} p_T^i {f}_{1T}^{\perp \; a}(x,p_T^2)$.
Here the focus will be on the $Q$ and $Q_T$ dependence, rather than on the $x$ and $z$ dependence. Therefore, we will assume that the TMDs as functions of $b_*$ are slowly varying functions of $b$ in the dominant $b$ region and consider them at a fixed value of $b_* \approx 0$ [\refcite{Boer:2013zca}]:
\beq
A_{UT}(x,z,Q_T) \approx \frac{x\, z^2\,\left(1-y+\frac{1}{2}y^2\right) \sum_{a,\bar a} e_a^2 f_{1T}^{\perp\prime \; a}(x;Q_0) D_1^a(z;Q_0)}{x\, z^2\, \left(1-y+\frac{1}{2}y^2\right) M Q_T  \sum_{b,\bar b} e_b^2 f_{1}^{b}(x;Q_0) D_1^b(z;Q_0)} {\cal A}(Q_T), 
\eeq
where 
\beq
{\cal A}(Q_T) \equiv M \frac{\int db\, b^2 \, J_1(bQ_T) \,\exp\left({-S_p(b_*,Q,Q_0)-S_{NP}(b,Q/Q_0)}\right)}{\int db\, b \, J_0(bQ_T) \, \exp\left({-S_p(b_*,Q,Q_0)-S_{NP}(b,Q/Q_0)}\right)} .
\eeq
The motivation for this approximation is that the same factor ${\cal A}(Q_T)$ will then appear in all asymmetries involving one $\boldsymbol{b}$-odd TMD, such as the Collins asymmetry, 
and not only in SIDIS, but also in  $e^+ e^- \to h_1 h_2 X$ and DY. It also permits a direct comparison to previous expressions in [\refcite{B01,B09}], 
allowing to estimate the impact of the new factorization on old results. Of course, the Sivers and Collins asymmetry will not evolve exactly in the same way, 
especially if one includes $x$ and $z$ dependence in $S_{NP}$, which is most relevant at low (HERMES and COMPASS) energies [\refcite{Anselmino:2012aa,APR}], but we expect that this approximation captures the dominant $Q$ and $Q_T$ behavior, especially for larger $Q$ of relevance to a future Electron-Ion Collider (EIC). 

In Ref.\ [\refcite{Boer:2013zca}] the factor ${\cal A}(Q_T)$ was studied numerically. It shows a peak at 
a $Q_T$ of a few GeV, which moves rather slowly towards larger values of $Q_T$ as $Q$ increases. 
${\cal A}(Q_T)$ decreases rapidly with increasing $Q$; on the peak it decreases as $1/Q^{0.7\pm 0.1}$. Here the uncertainty is estimated by varying $S_{NP}$ with an overall factor between 0.5 and 2. Although the fall-off is fast, it appears to be considerably slower than found in [\refcite{APR}], but there an integrated asymmetry was considered, which could explain the difference. 
The $Q$ behavior of the azimuthal asymmetries appears not to depend very much on the TMD factorization considered, e.g.\ [\refcite{CS81}] or [\refcite{Collins:2011zzd}], but the 
new TMD factorization is preferred because of its small $b$ ($<1/Q$) behavior, giving a much smaller contribution. To test the fall-off of the peak of the Sivers asymmetry requires a large $Q$ range, and hence an EIC, although the future very precise low-$Q$ data from Jefferson Lab 12 GeV will allow to extract 
$S_{NP}$ much more precisely and thereby help to reduce the uncertainty in the power of the fall-off. 

\subsection{Evolution of the double Collins effect}

The TMD fragmentation correlator for unpolarized final state hadrons is ($b=|\boldsymbol{b}|$):
\beq
\tilde{\Delta}(z,\boldsymbol{b}) = \frac{M}{4}\,\Biggl\{\tilde{D}_1(z,b^2)\, \frac{\not \! P}{M} + \left(\frac{\partial}{\partial b^2} \tilde{H}_1^\perp (z,b^2)\right) \, \frac{2 \not \! {\boldsymbol{b}} \not \! P}{M^2} \Biggl\}.
\eeq
The $\boldsymbol{b}$-odd second term parameterizes the Collins effect [\refcite{Collins:1992kk}]. 
In the process $e^+ e^- \to h_1 h_2 X$ a $\cos 2\phi$ asymmetry arises from the double Collins effect [\refcite{Boer:1997mf}]:  
\beq
\frac{d\sigma (e^+e^-\to h_1h_2X)}{dz_1 dz_2 d\Omega d^2{\boldsymbol q_T^{}}} \propto \left\{ 1 + \cos 2\phi_1 A(\boldsymbol{q}_T^{}) \right\},
\eeq
 \ba
{\rm where}\ A(Q_T) & \approx& \frac{\sum_a e_a^2 \sin^2 \theta \; H_1^{\perp (1) a}(z_1;Q_0) \; \overline H{}_1^{\perp (1) a}(z_2;Q_0)}{\sum_b e_b^2(1+\cos^2 \theta) \; D_1^b(z_1;Q_0) \; \overline D{}_1^b(z_2;Q_0)} \; {\cal A}(Q_T), \nonumber\\
{\rm and}\quad {\cal A}(Q_T) & \equiv & M^2 \frac{\int db\, b^3 \, J_2(bQ_T) \,\exp\left({-S_p(b_*,Q,Q_0)-S_{NP}(b,Q/Q_0)}\right)}{\int db\, b \, J_0(bQ_T) \, \exp\left({-S_p(b_*,Q,Q_0)-S_{NP}(b,Q/Q_0)}\right)}. \nonumber
\ea
Here it is assumed that the TMD FFs as function of $b_*$ are slowly varying. In this approximation the same overall factor 
${\cal A}(Q_T)$ appears also in the $\cos 2\phi$ asymmetries in SIDIS and DY.
\begin{figure}[htb]
\centerline{
\includegraphics[width=6.4 cm]{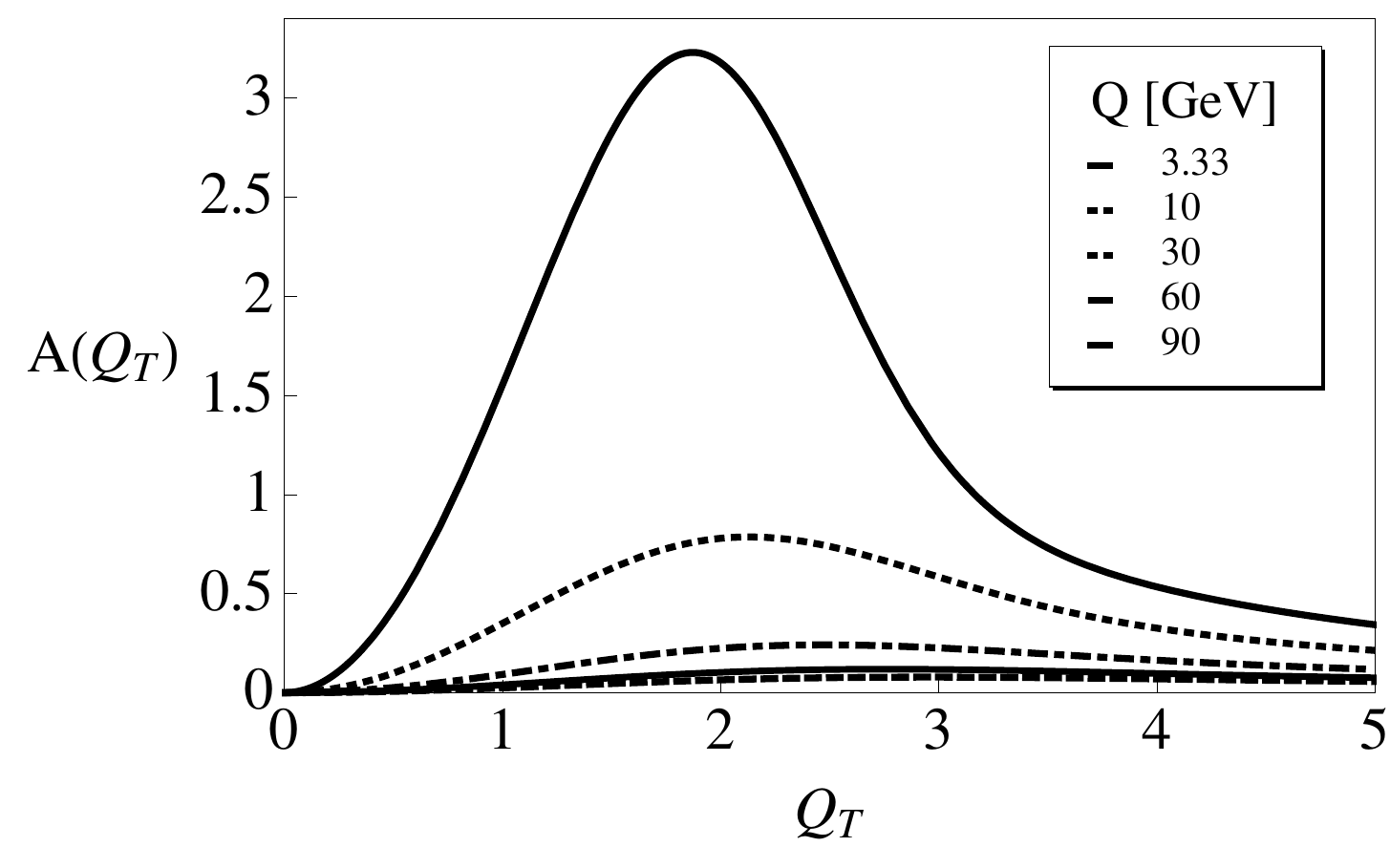}\hspace{1 mm}
\includegraphics[width=6.3 cm]{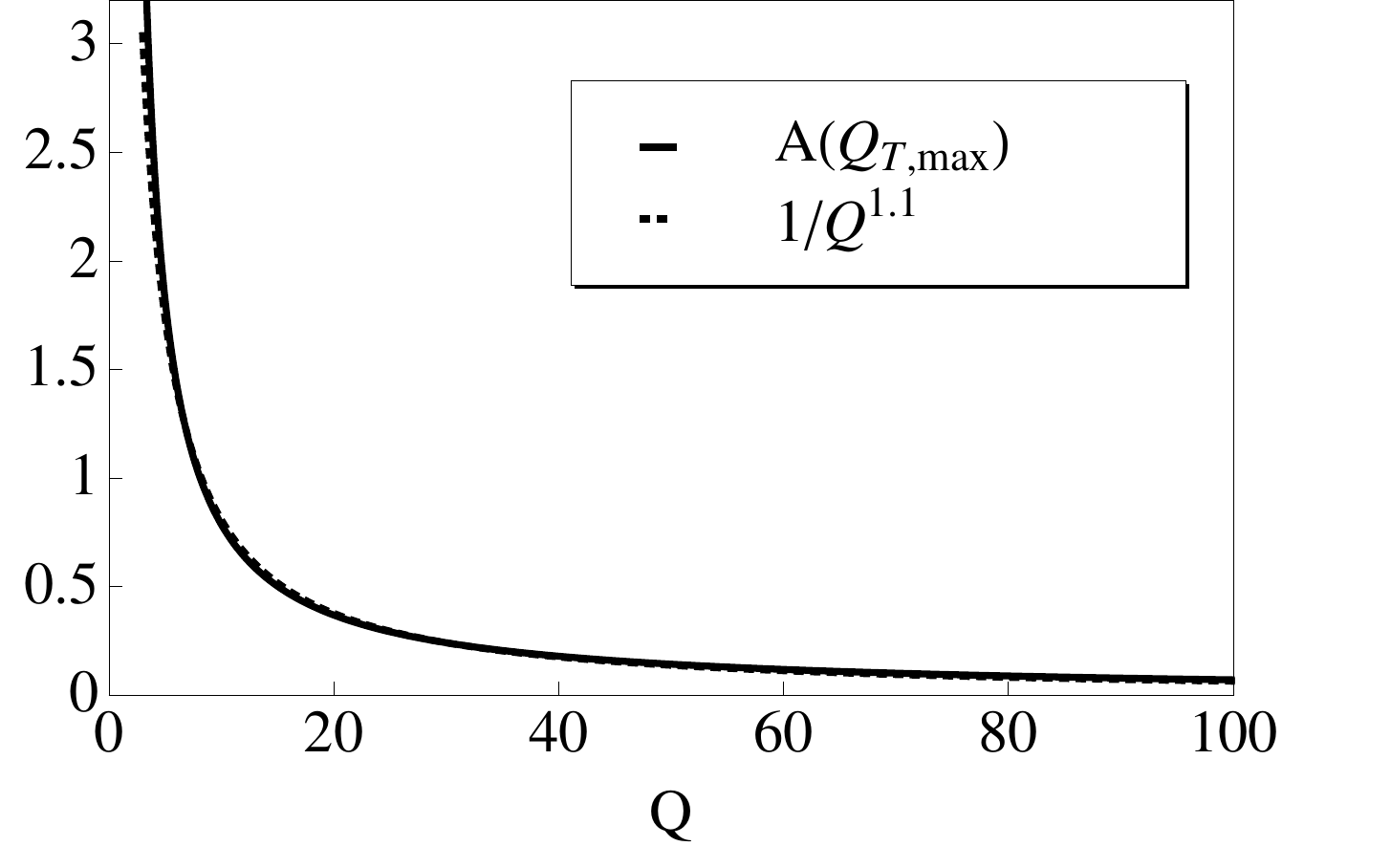}
}
\caption{\label{AQT}Left: the double Collins effect asymmetry factor ${\cal A}(Q_T)$ (in units of
$M^2$) for various $Q$. Right: ${\cal A}(Q_T)$ evaluated at the transverse momentum position $Q_{T,\max}$ of its maximum (solid line) plotted as function of $Q$  and compared to a line (dotted) with $1/Q^{1.1}$ fall-off.} 
\end{figure}
Fig.\ \ref{AQT} shows that there is considerable Sudakov suppression in the double Collins asymmetry, i.e.\ ${\cal A}(Q_T)$ falls off approximately as $1/Q^{1.1}$, which is consistent with the results of [\refcite{B01,B09}], which were based on the original TMD factorization by Collins and Soper from 1981 [\refcite{CS81}]. Experimentally the $\cos 2\phi$ asymmetry in $e^+ e^- \to h_1 h_2 X$ has been clearly observed in BELLE [\refcite{Abe:2005zx,Seidl:2008xc}] and BaBar [\refcite{Garzia}] data, but both are at the same scale. A similar measurement at the Beijing Electron Positron Collider could be very useful for a test of TMD evolution, see also  [\refcite{Sun:2013dya}].

\section{Comparison of high and low $Q_T$}

The TMD formalism applies for $Q_T^2 \ll Q^2$, which includes the intermediate region $\Lambda^2 \ll Q_T^2 \ll Q^2$, where one may also consider the 
low-$Q_T$ limit of the collinear factorization treatment valid at high $Q_T$ ($\Lambda^2 \ll Q_T^2$). Here we will discuss whether the two descriptions match or not, and whether this should be expected or not. 

If one considers the high-$Q_T$ limit of the low-$Q_T$ result, the small-$b$ region 
becomes dominant and the perturbative tail of the TMDs must be considered. 
The small-$b$ tail can be calculated within perturbation theory and 
leads to an integral over momentum fractions and mixing between quark and gluon distributions [\refcite{CS81,AR}]:
\[
\tilde{f}^a(x,\boldsymbol{b};\mu,\zeta_F) = \sum_{j=q,g} \int_x^1 \frac{d\hat{x}}{\hat{x}} \tilde{C}_{a/j}(x/\hat{x},b;\mu,g(\mu),\zeta_F) f^j(\hat{x};\mu) + {\cal O}\left((\Lambda_{\rm QCD} b)^n\right),
\]
for generic TMD $f$ (either pdf or FF) and some power $n$. 
For the unpolarized, angular independent part of the cross section, this perturbative expansion of the unpolarized TMDs ($\tilde{f}_1$ and $\tilde{D}_1$) 
yields an expression that precisely matches onto the leading logarithmic behavior of the high $Q_T$ expressions obtained in 
collinear factorization at fixed order in perturbative QCD (pQCD). Explicitly, to first order in $\alpha_s$ the low-$Q_T$ limit of the high-$Q_T$ expression is
\begin{align}
F_{UU, T}
&= \frac{1}{q_T^2}\, \frac{\alpha_s}{2\pi^2 z^2}
\sum_a x\ms e_a^2\,
\biggl[f_1^a(x)\,D_1^a(z)\,L\biggl( \frac{Q^2}{q_T^2} \biggr)
+ f_1^a(x)\, \bigl( D_1^a \otimes P_{qq}
                     + D_1^g \otimes P_{gq} \bigr)(z)
\nonumber \\
& \qquad\hspace{7em}
+ \bigl( P_{qq} \otimes f_1^a 
+ P_{qg} \otimes f_1^g \bigr)(x)\, D_1^a(z)
\biggr],
\label{low$Q_T$ofhigh$Q_T$}
\end{align}
up to power corrections in $q_T/Q$ and $M/q_T$. Here 
$L\left( Q^2/q_T^2\right) =
2 C_F \ln Q^2/q_T^2 - 3 C_F$ 
and $P_{ij}$ are the usual unpolarized splitting functions. 

At the same order the high-$Q_T$ limit of the low-$Q_T$ expression follows from inserting the perturbative tails of the 
unpolarized TMDs, which are given by
\begin{align}
f_1^{q}(x,p_T^2)
&= \frac{\alpha_s}{2 \pi^2}\,
   \frac{1}{\bm{p}_T^2}\,
   \biggl[\ms \frac{L(\eta^{-1})}{2}\, f_1^q(x)
     - C_F\ms f_1^q(x) 
     + \bigl(P_{qq} \otimes f_1^q + P_{qg} \otimes f_1^g\bigr)(x)
   \biggr] \,, \nonumber
\raisetag{-0.2em}
\\[0.4em]
D_1^{q}(z,k_T^2) &=
\frac{\alpha_s}{2 \pi^2}\,
\frac{1}{z^2\ms {\bm k}{}_T^2}\,
  \biggl[\ms \frac{L( \eta_{\smash{h}}^{-1} )}{2}\, D_1^q(z)
   - C_F D_1^q(z)
   + \bigl(D_1^q \otimes P_{qq} + D_1^g \otimes P_{gq}\bigr)(z)
  \biggr] \,, \nonumber
\end{align}
into the low-$Q_T$ expression $F_{UU ,T} = \mathcal{C}\left[ f_1 D_1 \right]$, which yields precisely the same answer as in Eq.\ (\ref{low$Q_T$ofhigh$Q_T$}) (for the standard definition of the convolution ${\cal C}[...]$ cf.\ Ref.\ [\refcite{BBDM}]).
This matching of the TMD factorization expression of the angular integrated cross section at low $Q_T$ to the fixed order pQCD collinear factorization 
expression at high $Q_T$ works to all orders in $\alpha_s$ and allows to obtain an expression that is valid over the whole $Q_T$ range, 
the well-known CSS resummation expression [\refcite{CSS-85}]. 

For the $\cos \phi$ and $\cos 2\phi$ dependences of the cross section this matching of high and low $Q_T$ expressions does not work [\refcite{Boer:2006eq,BBDM}], 
as we will now discuss.

The low-$Q_T$ limit of the high-$Q_T$ expression for the $\cos\phi$  dependence is
\begin{align}
F_{UU}^{\cos\phi_h}
&= - \frac{1}{Q\ms q_T}\, \frac{\alpha_s}{2\pi^2 z^2}
\sum_a x\ms e_a^2\,
\biggl[f_1^a(x)\,D_1^a(z)\,L\biggl( \frac{Q^2}{q_T^2} \biggr)
\nonumber \\
& \hspace{-1em} + f_1^a(x)\, \bigl( D_1^a \otimes P_{qq}'
                     + D_1^g \otimes P_{gq}'\bigr)(z)
+ \bigl( P_{qq}' \otimes f_1^a 
+ P_{qg}' \otimes f_1^g\bigr)(x)\, D_1^a(z)
\biggr],
\end{align}
The same logarithm as before appears, but the ``wrong" splitting functions, i.e.\ not 
those of the tails of the unpolarized TMDs. Therefore, CSS resummation
involving the standard angular-independent Sudakov factor cannot apply in this case [\refcite{Boer:2006eq}].

In the low-$Q_T$ TMD region a $\cos \phi$ asymmetry is generated at twist-3, involving the 
twist-3 TMD $f^\perp$ and TMD FF $D^\perp$ (which includes the Cahn effect) [\refcite{Levelt:1994np}]:
\begin{equation} 
F_{UU}^{\cos\phi_h}
= - \frac{2}{Q}\,\mathcal{C}\biggl[
   \hat{\bm{h}}\cdott \bm{k}_T^{}\,f_1 \frac{\tilde{D}^{\perp }}{z}
   + \hat{\bm{h}}\cdott \bm{p}_T\,x f^{\perp } D_1
   \biggr] + {\rm chiral}\mbox{-}{\rm odd}\ {\rm terms}.
\end{equation}
It so happens that the ``wrong'' splitting functions correspond to those 
of the perturbative tails of $f^\perp$ and $D^\perp$ [\refcite{BBDM}] (the tails of the chiral-odd functions are suppressed):
\begin{align}
x f^{\perp q}(x,p_T^2)
&= \frac{\alpha_s}{2 \pi^2}\,
   \frac{1}{2 \bm{p}_T^2}\,
   \biggl[\ms \frac{L(\eta^{-1})}{2}\, f_1^q(x)
     + \bigl(P'_{qq} \otimes f_1^q + P'_{qg} \otimes f_1^g \bigr)(x)
   \biggr] \,, \nonumber
\raisetag{-0.2em}
\\[0.4em]
\frac{\tilde{D}^{\perp q}(z,k_T^2)}{z} &=
- \frac{\alpha_s}{2 \pi^2}\,
  \frac{1}{2 z^2\ms {\bm k}{}_T^2}\,
  \biggl[\ms \frac{L( \eta_{\smash{h}}^{-1} )}{2}\, D_1^q(z)
   - 2 C_F D_1^q(z)
   + \bigl(D_1^q \otimes P'_{qq} + D_1^g \otimes P'_{gq}\bigr)(z)
  \biggr] \,. \nonumber
\end{align}
Inserting this in the ``naive'' (unproven) twist-3 TMD factorized expression yields
the high-$Q_T$ result except for an extra term $-2 C_F f_1^a(x)
\,D_1^a(z)$. In other words, assuming TMD factorization beyond leading twist, the $\cos\phi$ term 
at low $Q_T$ almost matched onto the collinear factorization result at high $Q_T$. 
This hints at a possible modified TMD factorization at twist-3, at least for the completely 
unpolarized case.

The $\cos 2\phi$ asymmetry as function of $Q_T$ has dominant high and low $Q_T$ contributions from 
different origins. At low $Q_T$ the asymmetry is proportional to $h_1^\perp H_1^\perp$, which has a $M^2/Q_T^2$ suppressed high-$Q_T$ tail. 
At high $Q_T$ it is proportional to  $f_1 D_1$, which in turn is $Q_T^2/Q^2$ suppressed at low $Q_T$.  
These two contributions do not and need not match in the intermediate region $\Lambda^2 \ll Q_T^2 \ll Q^2$. 
To power correction accuracy they can simply be added [\refcite{BBDM}]. As we have seen, the low-$Q_T$ part falls off approximately as $1/Q$, which is slower than the $1/Q^2$ fall-off of the high-$Q_T$ asymmetry expression. This would have posed a problem if one had to match them. 

For the calculation of the high-$Q_T$ limit of the TMD expression for the Sivers asymmetry one needs the following perturbative tails: 
\[
f_1(x,\boldsymbol{p}_T^2) \stackrel{\boldsymbol{p}_T^2 \gg M^2}{\sim}  \! \alpha_s\,\frac{1}{\boldsymbol{p}_T^2} \, \left(K \otimes f_1\right) (x), \quad  
f_{1T}^\perp(x,\boldsymbol{p}_T^2) \stackrel{\boldsymbol{p}_T^2 \gg M^2}{\sim} \! \alpha_s\, \frac{M^2}{\boldsymbol{p}_T^4} \, \left(K' \otimes T_F\right) (x).
\]
It shows that the twist-3 Qiu-Sterman function $T_F$ [\refcite{Qiu:1991pp}],  with operator definition 
\beq
T_F(x,x) \stackrel{A^+ =0}{\propto}  \text{F.T.} \, \langle P|\; {\overline \psi} (0) \; {\int \! d\eta^- \; F^{+\alpha}(\eta^-)}\; \gamma^+ \, \psi(\xi^-) \;|P \rangle,
\eeq
determines the large-$p_T$ behavior of the Sivers TMD. Insertion of the perturbative tails in the TMD expression yields precisely the high-$Q_T$ result 
[\refcite{Ji:2006ub,Koike:2007dg}]. This matching holds to all orders and a CSS resummation expression for the Sivers asymmetry 
has been derived in [\refcite{Kang:2011mr}]. The $Q$ behavior of the high-$Q_T$ expression is not straightforward. The evolution of $T_F(x,x)$ is known 
[\refcite{Kang:2008ey}-\refcite{Braun:2009mi}], but is non-autonomous,  
depending on $T_F(x,y)$ for 
$x\neq y$. It only becomes autonomous in the large-$x$ limit, which for the non-singlet case implies
$T_F(x,x;Q^2)/f_1(x,Q^2) \sim \left( \alpha_s(Q)/\alpha_s(\mu_0) \right)^{2N_c/b_0}$ [\refcite{Braun:2009mi}], i.e.\
$T_F$ evolves logarithmically with $Q^2$, but faster than $f_1$. 

\section{Bessel-weighted asymmetries}

Transverse momentum weighted asymmetries were first considered, because azimuthal asymmetries in differential cross sections are convolution expressions of TMDs that appear in 
different processes in different ways, whereas weighting projects out ``portable'' functions [\refcite{Kotzinian:1995cz}].
It was also observed that specific weighted asymmetries are insensitive to Sudakov suppression (the $1/Q^\alpha$ behavior of azimuthal asymmetries 
at low $Q_T$) [\refcite{B01}]. E.g.\ for single spin asymmetries that applies to the weighted integral 
$\int d^2 \boldsymbol{q}_T^{} \, {\boldsymbol{q}_T^{i}} \, d\sigma/d^2 \boldsymbol{q}_T^{} 
\rightarrow \int d^2 \boldsymbol{b} \, \delta^2(\boldsymbol{b}) \, \partial\tilde{W}(b)/\partial b^i$.
The latter integral is only nonzero for $\boldsymbol{b}$-odd TMDs, such as the Sivers function, yielding an asymmetry proportional to $\exp(-S(b=0)) \; {{f}_{1T}^{\perp\prime}(x;Q_0)\; {D}_1(z;Q_0)}$, where $S(b=0)=0$. The only remaining $Q$ dependence is through $H(Q;\alpha_s(Q))$. Consequently, the weighted Sivers asymmetry must evolve logarithmically, which is consistent with the fact that the weighted Sivers function is proportional to the Qiu-Sterman function [\refcite{BMP}].

This weighting with powers of $\boldsymbol{q}_T$ assumes that first of all, the integrals converge, and that, secondly, integrals over TMD expressions (without $Y$ term) are fine to begin with. 
For example, for the $\cos 2\phi$ asymmetry 
the appropriate (conventional) weighting 
would be with $\boldsymbol{q}_T^2$. 
Unfortunately the $\boldsymbol{q}_T^2$-weighted integral is sensitive mainly to the perturbative high-$Q_T$ part of the asymmetry 
($\sim Q_T^2/Q^2$). 

To by-pass these tricky issues that both concern the perturbative tails of the asymmetries, 
one can consider Bessel weighting instead [\refcite{Boer:2011xd}], where one replaces powers of the observed transverse momentum in the conventional weights
$|\boldsymbol{q}_T|^n$ by $J_n(  |\boldsymbol{q}_T| {\cal B}_T)\, n! \left(2/{\cal B}_T\right)^n$. 
If ${\cal B}_T$ is not too small, the TMD region should dominate and the $Y$ contribution suppressed. 
In the limit ${\cal B}_T \to 0$ conventional weights are retrieved, the $Y$ term becomes very important 
and divergences may arise.  

Furthermore, Bessel weighting has the advantage of allowing lattice QCD evaluations of TMDs, both T-even and T-odd ones. 
Consider the average transverse momentum shift orthogonal to a given transverse polarization --the Sivers shift--:
\[
\langle p_y(x) \rangle_{TU}  = \left. \frac{ \int d^2 {\boldsymbol p}_T\, {\boldsymbol p}_y  \ \Phi^{[\gamma^+]}(x,{\boldsymbol p}_T;\zeta_F,\mu^2) }{ 
\int d^2 {\boldsymbol p}_T\, \Phi^{[\gamma^+]}(x,{\boldsymbol p}_T;\zeta_F,\mu^2) } \right|_{S^\pm=0,{\boldsymbol S}_T=(1,0)} =  M \frac{ f_{1T}^{\perp(1)}(x;\zeta_F,\mu^2) }{ f_1^{(0)}(x;\zeta_F,\mu^2)} ,
\]
and its Bessel-weighted analogue (for details and definitions cf.\ [\refcite{Boer:2011xd}]): 
\begin{eqnarray}
\langle p_y(x) \rangle_{TU}^{{\cal B}_T} &  = & \left. \frac{ \int d |{\boldsymbol p}_T|\, |{\boldsymbol p}_T| \int d\phi_p \frac{2J_1(|{\boldsymbol p}_T|{\cal B}_T)}{{\cal B}_T}\, 
\sin(\phi_p-\phi_S)\, \Phi^{[\gamma^+]}(x,{\boldsymbol p}_T;\zeta_F,\mu^2) }{ \int d |{\boldsymbol p}_T|\,|{\boldsymbol p}_T| \int d\phi_p J_0(|{\boldsymbol p}_T|{\cal B}_T)) \, 
\Phi^{[\gamma^+]}(x,{\boldsymbol p}_T;\zeta_F,\mu^2) } \right|_{|{\boldsymbol S}_T|=1} \nonumber\\ 
&  =  & M \frac{ \tilde{f}_{1T}^{\perp(1)}(x,{\cal B}_T;\zeta_F,\mu^2) }{ \tilde{f}_1^{(0)}(x,{\cal B}_T;\zeta_F,\mu^2)} .
\end{eqnarray}
After taking Mellin moments, the Bessel-weighted Sivers shift yields a well-defined quantity $\langle k_T \times S_T\rangle (n,{\cal B}_T)$, that can be evaluated on the lattice. This approach has led to the first `first-principle' demonstration in QCD that the Sivers function including proper gauge links is nonzero [\refcite{Musch:2011er}]. It also clearly corroborates the sign change relation $f_{1T}^{\perp [{\rm SIDIS}]} = - f_{1T}^{\perp [{\rm DY}]}$ [\refcite{Collins:2002kn}]. 
TMD evolution is very important for the experimental test of this sign relation, as the Sivers function may have a scale dependent node as a function $x$ and/or $k_T$ [\refcite{Boer:2011fx,Kang:2011hk}]. As a further complication, 
any node can be at different places for different flavors, although one expects $f_{1T}^{\perp u}(x,k_T^2) = - f_{1T}^{\perp d}(x,k_T^2) + {\cal O}(1/N_c)$ [\refcite{Pobylitsa:2003ty,Drago:2005gz}].
Some model calculations show a node, but not for all flavors (for $u$-quarks see e.g.\ [\refcite{Bacchetta:2008af}] and for $d$-quarks [\refcite{Lu:2004au,Courtoy:2008vi}]). 

\section{Summary}
From the numerical study of the TMD evolution of azimuthal asymmetries we observe that the new TMD factorization [\refcite{Collins:2011zzd}] does not change much the power of the $1/Q^\alpha$ fall-off of azimuthal asymmetries compared to earlier TMD factorization expressions considered in the literature. Partially integrated asymmetries may fall off faster than the peak. The dependence on the non-perturbative Sudakov factor and its correct form is relevant at all experimentally accessible scales and need to be investigated further. 

The $Q$ dependences of the high and low $Q_T$ contributions generally differ.
Matches and mismatches of high and low $Q_T$ contributions occur: the Sivers asymmetry matches; the $\cos \phi$ asymmetry almost matches, but requires modified TMD factorization at twist-3; the $\cos 2\phi$ asymmetry does not match, but is understood: to power correction accuracy both contributions can be added, without double counting. 

Bessel weighting allows to project out ÒportableÓ functions, to consider convergent integrals over all $Q_T$, to emphasize the TMD region and suppress the high-$Q_T$ $Y$-term contribution, and last but not least, allows for lattice QCD evaluations of T-even and T-odd TMDs including proper gauge links. 

\section*{Acknowledgments}
I thank John Collins, Ted Rogers, Werner Vogelsang, and Feng Yuan for useful discussions, and Alessandro Bacchetta, Markus Diehl, Leonard Gamberg, Piet Mulders, Bernhard Musch, and Alexei Prokudin for pleasant collaboration on parts of the presented work. 


\end{document}